\documentclass[12pt]{article}
\usepackage{times}
\usepackage{geometry}
\usepackage[utf8]{inputenc}
\usepackage{enumitem,amssymb}
\usepackage{ragged2e}
\usepackage[bookmarksdepth=3,linktoc=all,colorlinks=true,urlcolor=blue,linkcolor=blue,citecolor=black]{hyperref}
\usepackage{natbib}
\usepackage{wrapfig, caption}
\usepackage{graphicx}
\usepackage{xcolor}

\usepackage{fancyhdr}
\newlist{thematic}{itemize}{8}
\setlist[thematic]{label=$\square$}
\usepackage{pifont}

\def\iso#1#2{\mbox{${}^{#2}{\rm #1}$}}
\def\mn5#1{\iso{Mn}{5#1}}
\def\fe6#1{\iso{Fe}{6#1}}
\def\pu24#1{\iso{Pu}{24#1}}

\def\pfrac#1#2{\left( \frac{#1}{#2} \right)}

\def\DSN{D_{\rm SN}}

\begin{document}
\thispagestyle{empty}
\Large
\centerline{Solar and Space Physics 2024 Decadal Survey White Paper}
\bigskip

\huge
\centerline{Near-Earth Supernovae in the Past 10 Myr:}
\centerline{Implications for the Heliosphere}
\bigskip \bigskip

\normalsize
\noindent \textbf{Primary Category:} Basic Research--Emerging Opportunities:  interstellar medium \\
\textbf{Secondary Category:} Outer Heliosphere 

\bigskip

\noindent
\textbf{Authors:} \\
\indent Jesse A.~Miller,  University of Illinois and Boston University  \\
\indent Brian D.~Fields,  University of Illinois 

\bigskip

\noindent
\textbf{Co-authors:} 
\linebreak
\indent Thomas Y. Chen, Columbia University \\
\indent John Ellis, Physics Dep't, King's College London, UK and CERN, Geneva, Switzerland \\
\indent Adrienne F.~Ertel, University of Illinois \\
\indent Jerry W.~Manweiler, Fundamental Technologies, LLC \\
\indent Merav Opher, Boston University \\
\indent Elena Provornikova, Applied Physics Laboratory, Johns Hopkins University \\
\indent Jonathan D.~Slavin, Center for Astrophysics $|$ Harvard \& Smithsonian  \\
\indent Justyna Sok\'o\l, Southwest Research Institute \\
\indent Veerle Sterken, ETH Zürich, Zürich, Switzerland \\
\indent Rebecca Surman, University of Notre Dame \\
\indent Xilu Wang, Institute of High Energy Physics, Chinese Academy of Sciences

\bigskip
  
\noindent
\textbf{Synopsis:} \linebreak
We live in a dynamic star-forming galaxy, and as a result the solar neighborhood changes constantly.  One of the most spectacular events that inevitably occurs is the explosion of a supernova near the Earth.  Remarkably, a wealth of evidence now shows that at least two supernovae occurred during the past 10 Myr within $\sim 100$ pc of Earth.  We summarize these data, most notably the detection of live radioactive \fe60 in the deep sea, Antarctic snow, and the Moon.  Analytic estimates and hydrodynamic simulations show that heliospheric response was dramatic, with the heliopause pushed to $\sim 20$ au. 

These supernovae and their impact on the solar system have wide implications beyond heliophysics, including the possibility of damage to Earth's biosphere, and open a new tool to probe supernova element production and dust evolution.
The wide scope of these implications illustrates the close and complementary relationship between heliophysics and other areas of science, and provides a compelling case for a cross-cutting, interdisciplinary research program.
This white paper advocates for strengthening cross-disciplinary research on nearby supernova impacts on the heliosphere, interstellar dust, and cosmic rays. We urge for support of investigation via theoretical work, direct exploration of the outer heliosphere and very  local interstellar medium, and study of extrasolar astrospheres.

\pagebreak
\setcounter{page}{1}

\section*{Evidence For Nearby Supernovae in the Past 10 Myr}

The environment surrounding the Sun changes constantly in our journey around the Milky Way, and the heliosphere constantly evolves in response.
\citet{muller_heliospheric_2006, muller_heliosphere_2009} demonstrate how the heliosphere's size depends greatly on its local Galactic environment,
largely depending on the density, velocity, and ionization state of the surrounding medium.
Recently, \citet{opher_climate_2022} simulated the effect of the Sun's passage through a dense, cold cloud, showing how the heliopause would shrink to a mere 0.22 au and expose the Earth to interstellar material.
In this white paper, we focus on the other most dramatic disturbance to the heliosphere:  the explosion of a near-Earth supernova.

Supernova explosions mark the deaths of massive stars \citep{Branch2017}, and launch powerful blast waves that sculpt the interstellar medium (ISM).
Nuclear reactions deep in the star before and during its death forge heavy elements, principally species from carbon through the iron peak.  The explosion launches these freshly synthesized elements into the surroundings.

How would we know if a supernova exploded near Earth in the geologic past? If the explosion is close enough, supernova debris is delivered to Earth and literally rains down, accumulating in natural archives such as deep-sea sediments. To confirm the supernova origin of such debris, one must look for live (undecayed) radioactivity from species too short-lived to have survived from the birth of the Earth 4.6 Gyr ago~\citep{Ellis:1995qb}.  Remarkably, such a signal has been found, in the form of live \fe60 (half-life 2.6 Myr) and \pu244 (half-life 81 Myr).

\begin{figure}[htb]
    \centering
    \includegraphics[width=0.9\textwidth]{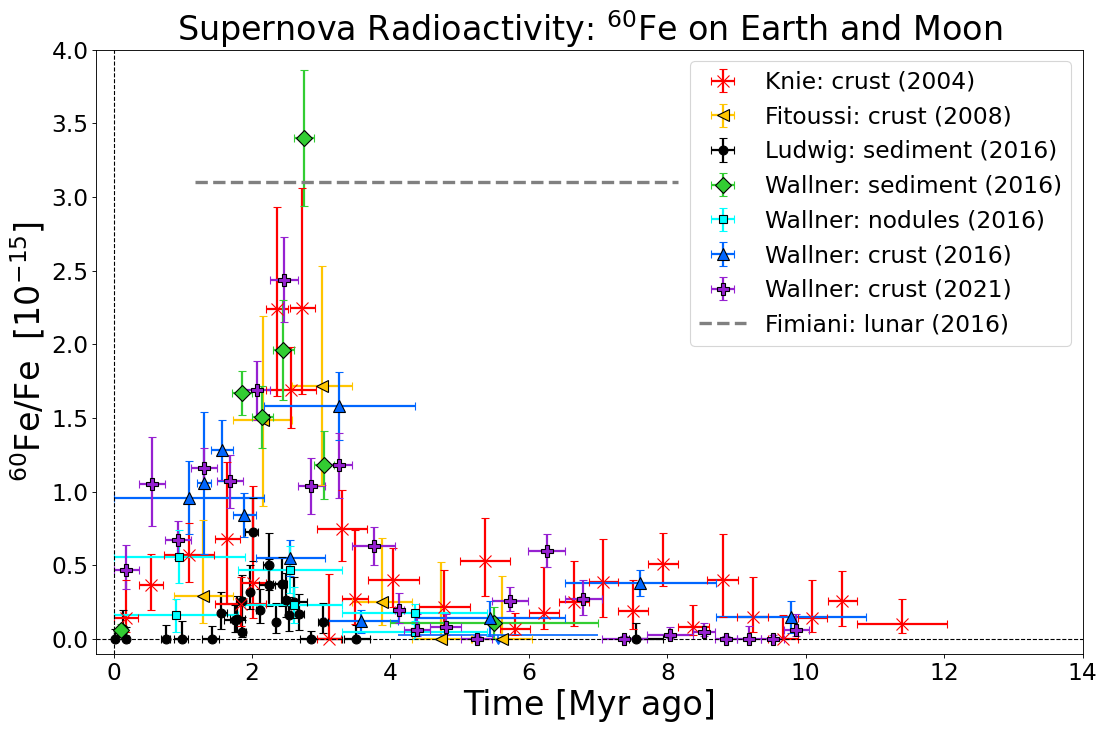}
    \captionsetup{width=0.9\textwidth}
    \caption{Evidence for recent near-Earth supernovae:  compilation of radioactive \fe60 measurements in the deep sea and lunar regolith, from \citet{ertel_supernova_2022}. 
    All measurements show a pulse 2--3 Myr ago, and recent data confirm a pulse 6--7 Myr ago. Nonuniform fallout and uptake on Earth are reflected in \fe60 abundance variations.
    }
    \label{fig:data}
\end{figure}

Fig.~\ref{fig:data} shows the profile of \fe60 recovered from deep-sea and lunar samples analyzed by several groups \citep{knie_ams_1999, knie_60fe_2004, fitoussi_search_2008, ludwig_time-resolved_2016, wallner_recent_2016, fimiani_interstellar_2016, wallner_60fe_2021}.
All measurements agree on the large peak $\sim 3$ Myr ago. Furthermore, \citet{wallner_60fe_2021} indicates the presence of a preceding supernova that exploded around 8 Myr ago, and measured \pu244 from periods around these peaks.

The \fe60 found in these terrestrial samples came from a core-collapse or electron-capture supernova, as other extrasolar options (e.g., thermonuclear supernovae or kilonovae) cannot produce sufficient \fe60 without being close enough to also inflict major biological damage \citep{fry_astrophysical_2015}. While super-asymptotic giant branch (AGB) stars expel \fe60 in their winds, their velocity is likely too slow to reach the Earth. Based on \fe60 production rates in supernovae, \citet{fry_astrophysical_2015} estimated the distance to the recent event 3 Myr ago as $D_\mathrm{SN} \sim 60-130$ pc.

Further evidence of \fe60 has been found in freshly-deposited ($< 20$ yr) Antarctic snow \citep{koll_interstellar_2019},
indicating that {\em supernova deris still falls upon the Earth}, though at a reduced flux.
Besides \fe60, other isotopes have been identified as potential tracers of supernova activity. Of these, \pu244 arrived concurrently with \fe60 \citep{wallner_60fe_2021}, and a claim of elevated \mn53 has been made \citep{korschinek_supernova-produced_2020}.

Beyond these radioisotope deposits, cosmic ray measurements suggest recent nearby supernovae.  
\citet{binns_observation_2016} measured \fe60 in cosmic rays and inferred it must have been present at the source, thus pointing to a recent and nearby event.  Measurements of anomalous Fe cosmic rays at low energies \citep{Boschini2021} and high-energy positrons and antiprotons \citep{savchenko_imprint_2015, kachelries_signatures_2015, kachelries_cosmic_2018}
could also  point to recent local sources.

Core-collapse supernovae are the end state of massive star evolution. Massive stars are overwhelmingly born in clusters, and are usually in binaries, often with other massive stars \citep{Duchene2013,Motte2018}.  Thus, we expect nearby supernovae are likely to occur as multiple events, and likely should be associated with massive star clusters.  The \citet{wallner_60fe_2021} confirmation of a distinct second \fe60 pulse is thus consistent with expectations for massive star formation.

Indeed, prior to the \fe60 discoveries, evidence for recent nearby supernovae was literally all around us, in the form of the Local Bubble--a cavity of low-density, high-temperature plasma in which the solar system is embedded \citep{frisch_interstellar_2011}.   The Local Bubble has been mapped in gas via UV metal lines \citep{Gry2014}, optical lines \citep{Welsh_etal_2010}
and diffuse X-rays \citep{Galeazzi2014,Snowden2015}; it has also been mapped in dust via observations of diffuse interstellar bands \citep{Farhang2019} and interstellar reddening \citep{Lallement2019}.
The Local Bubble is irregular in shape, and the boundaries are not precise, but it extends 50 pc or more in all directions. Models of this large, hot cavity have invoked two or more supernovae \citep[e.g.,][]{smith_multiple_2001, frisch_interstellar_2011, breitschwerdt_locations_2016, schulreich_numerical_2017, frisch_effect_2017,Zucker2022}.  
These supernovae likely arose in massive star clusters that may have stars remaining to this day \citep{Maiz-Apellaniz_2001}.  These are also candidates for the origin of the \fe60, and include the Scorpius-Centaurus Association at $\sim 130 \ \rm pc$ \citep{Benitez2002} and Tucanae-Horologium
at $\sim 50 \ \rm pc$ \citep{Mamajek2016}.  There have  been attempts made to identify the \fe60 neutron star with a nearby pulsar, and to associate it kinematically with a potential companion runaway star by tracing back their paths over time to the Sco-Cen Association \citep{neuhauser_nearby_2020}.

The Local Bubble thus sets the gross features of the interstellar environment around the Sun.
In short, we reside in a so-called ``superbubble'' created by up to $\sim 20$ supernovae \citep{schulreich_numerical_2017}. The very local interstellar medium that controls the heliosphere should be understood in this larger context. This means that direct probes of interstellar space immediately  beyond the Sun also open a new window into a superbubble interior.  Interestingly, the model of \citet{Zucker2022} suggests that the solar system entered the Local Bubble $\sim 5 \ \rm Myr$ ago.  This date is uncertain, but is intriguing in light of the earlier \fe60 pulse dating to $\gtrsim 8 \ \rm Myr$ ago; this may probe our entry into the Local Bubble.  
Moreover, the radioisotope data as well as the astronomical data on the Local Bubble together  reveal the recent nearby supernova history of our interstellar neighborhood, and consequently, the extreme environments our solar system has travelled through.

\section*{Implications for the Heliosphere}

A near-Earth supernova would dramatically alter the heliosphere for many thousands of years.  The full implications of such an event are wide-ranging and many remain to be explored.  But the major dynamical effects have begun to come into focus.

\begin{wrapfigure}[24]{r}{0.48\textwidth}
    \vspace*{-0.5cm}
    \includegraphics[width=0.48\textwidth]{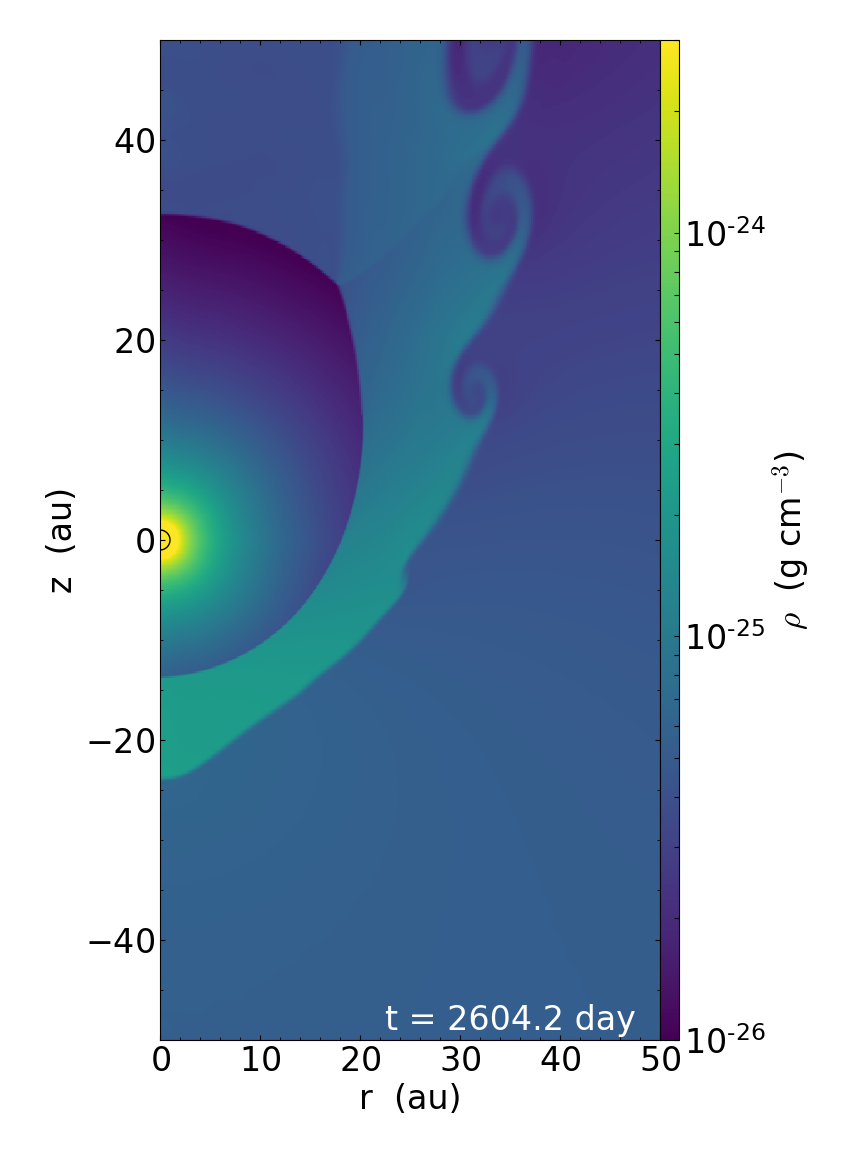}
    \vspace*{-1.2cm}
    \captionsetup{width=0.45\textwidth}
    \caption{A simulation of the heliosphere being compressed by a supernova 63 pc distant, shown in \citet{miller_heliospheric_2022}. The blast wave compresses the heliosphere to $\sim 20$ au, exposing much of the outer solar system to the blast.}
    \label{fig:63pc_sim}
\end{wrapfigure}

A supernova blast creates a powerful forward shock moving at speeds $v_{\rm SN} \sim 200-1000 \ \rm km/s$ depending on the distance to the explosion and the density of the surrounding medium.
The blast is supersonic, so that ram pressure dominates over thermal pressure. At a distance $\DSN$ the pressure is roughly
\begin{eqnarray}
\label{eq:P_SN}
    P_{\rm SN} & \approx & \frac{E_{\rm SN}}{\DSN^3} \\ 
    & \approx & 3 \times 10^{-10} \ {\rm dyne/cm^2}
    \ \pfrac{50 \ \rm pc}{\DSN}^3 \\
    & \approx & P_{\rm SW}(10 \ {\rm au}) \, , \label{eq:P_10au}
\end{eqnarray}
where we have used the canonical supernova explosion energy (excluding neutrinos) of $E_{\rm SN} = 10^{51} \ \rm erg$.
This result holds when the supernova remnant (SNR) is in the Sedov-Taylor phase, as is appropriate for our problem.
Conveniently, in this phase the pressure is {\em independent} of the density of the medium surrounding the explosion:  lower densities lead to higher blast speeds that combine to give the same ram pressure.

Eqn.~(\ref{eq:P_SN}) shows that the supernova pressure is 
$> 100$ times the present-day pressure of the Local ISM,
for supernova distances implied by the \fe60 detections.
Thus we expect dramatic effects on the heliosphere
accompanied these events.  
Indeed, eqn.~(\ref{eq:P_10au}) shows that in pressure balance, we expect the blast to penetrate to around 10 au!

To date, the only studies of nearby supernova blasts impacting the heliosphere are~\cite{fields_supernova_2008} and \cite{miller_heliospheric_2022}. These initial simulations explored a wide range of possible supernova blast distances and interstellar distances in order to characterize the compression of the heliosphere.
Fig.~\ref{fig:63pc_sim} shows an example of these simulations for a supernova 63 pc away, a plausible distance for the recent supernovae.
The supernova blast has speed 541 km/s
and flows from the bottom to top of the figure.
We see that the arrival of such a supernova creates a heliosphere with the same structures as the modern version \citep{baranov_model_1993}---termination shock, heliopause, and bow shock---but extremely compressed.
In this case, the supernova material reaches $\sim 20$ au, where we see the heliopause.
\citet{miller_heliospheric_2022} showed that the basic geometry and the locations of the features are similar whether the blast is directed towards the Sun's equator or the poles; we thus expect that the relative orientation of the Sun and supernova is not a major factor in the compression.

Both \citet{fields_supernova_2008} and \citet{miller_heliospheric_2022} found that force balance is the controlling factor determining the heliospheric compression in these simulations. That is, the closest approach of the supernova is set by the balance of the blast pressure with that of the solar wind.  Along the symmetry axis, this gives a  stagnation radius
\begin{equation}
    \label{eqn:rstag}
    r_{\rm stag} = 24 {\rm\ au} \pfrac{\DSN}{100\ {\rm pc}}^{3/2} \pfrac{E_{\rm SN}}{10^{51}\ {\rm erg}}^{-1/2} \, ,
\end{equation}
in good agreement with the numerical simulation results.
As in eqn.~(\ref{eq:P_SN}), $\DSN$ is the distance to the supernova, and $E_{\rm SN}$ is the supernova explosion energy.
For a fixed supernova explosion energy, therefore, the most important parameter controlling the heliospheric compression is the distance to the supernova.

\begin{wrapfigure}[21]{r}{0.7\textwidth}
    \includegraphics[width=0.68\textwidth]{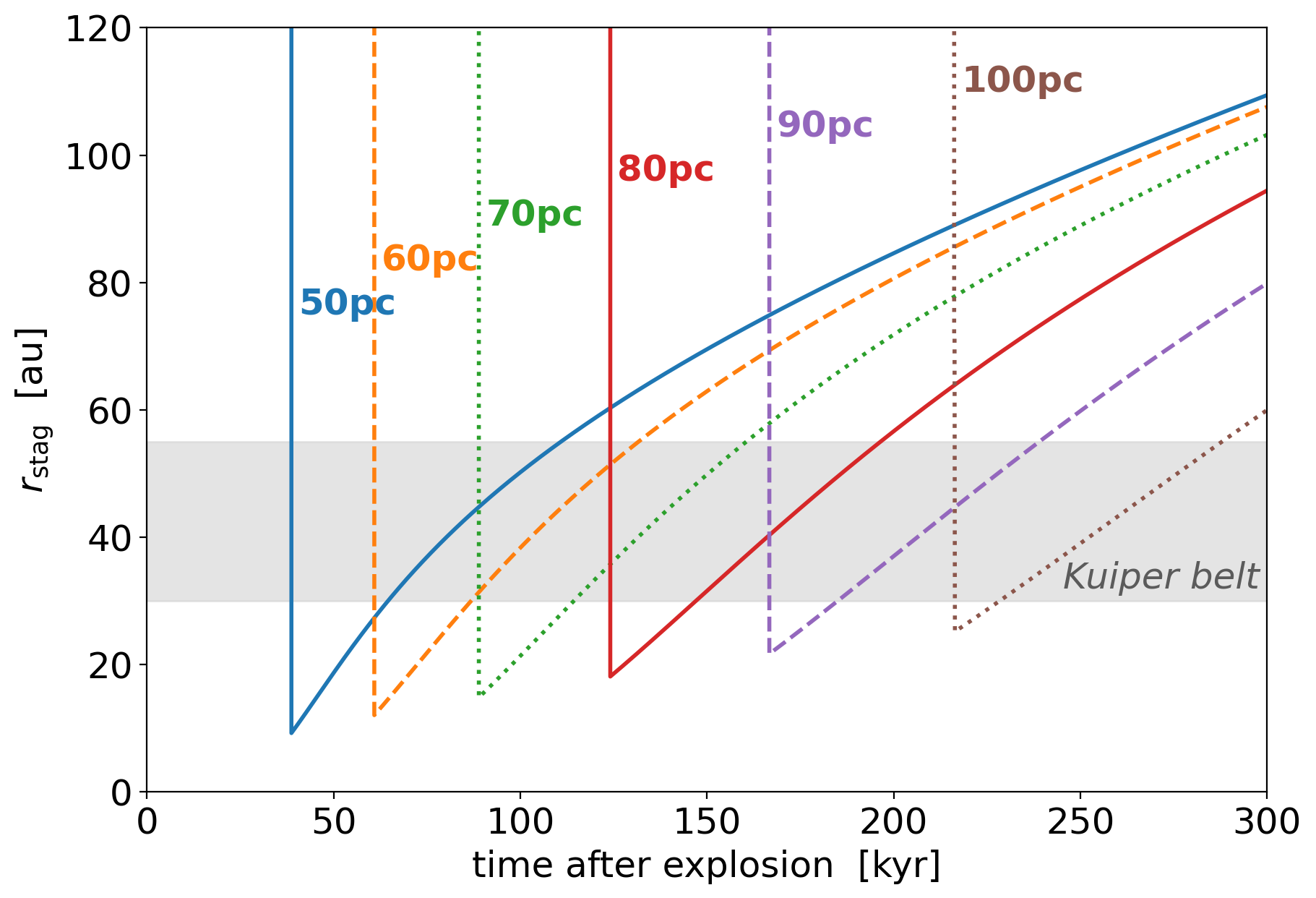}
    \vspace{-0.4cm}
    \centering
    \captionsetup{width=0.65\textwidth}
    \caption{Location of the stagnation radius (ram pressure balance) over time as a supernova remnant evolves, from \citet{miller_heliospheric_2022}. The distances to the supernovae are labelled, and for reference the Kuiper belt is indicated by the gray region.}
    \label{fig:Rstag}
\end{wrapfigure}

The scaling in eqn.~(\ref{eqn:rstag}) holds for purely hydrodynamic simulations.
These simulations were simplified, imposing 2D axisymmetry, adopting the hydrodynamics of a one-component fully ionized and unmagetized plasma, and using a non-varying solar wind.
Further study is needed to illuminate the importance of additional physical effects such as magnetic fields, charge exchange, pickup ions and energetic neutral atoms \citep[e.g.,][]{sokol_interstellar_2019, Sokol2022}, and cosmic ray and dust propagation \citep[see, e.g.,][for a review]{zank_interaction_1999}. These effects are significant for the present-day heliosphere, but are wholly unstudied under a supernova blast regime, which could also have novel effects depending on the density of the impacting gas, such as radiative cooling. 

While the initial supernova shock front sweeps rapidly across the solar system, the whole supernova remnant compresses the heliosphere for up to $\sim$ 100 kyr until it weakens enough for the heliosphere to rebound.  Using pressure balance, and following the propagation of a Sedov-Taylor blast, we can calculate the retreat of the stagnation radius after the supernova forward shock engulfs the solar system.  Fig.~\ref{fig:Rstag} shows the size of the heliosphere over time for supernovae at distances from 50-100 pc.  We see that a significant fraction of the solar system is directly exposed to the blast wave for all plausible supernova distances. Even for a supernova 100 pc away, the entire Kuiper belt is exposed.

This analysis assumes the Sun is at rest relative to the supernova remnant.
However, including solar motion has significant effects on the gas profile and \fe60 deposition.
\citet{chaikin_simulations_2022} shows how the relative motion of the Sun elongates the \fe60 deposition timescale to a few Myr, even without dust dynamics analyzed in \citet{fry_magnetic_2020}. This work demonstrates how the gradual rebounding of the heliosphere will not be as smooth as the idealized case in Fig.~\ref{fig:Rstag}.

On the other hand, even for reasonably close supernovae (50 pc), interstellar material does not reach Earth's orbit. Therefore, the \fe60 observed in deep sea samples could not have arrived as a plasma, but must have come in the form of dust in order to not be deflected by the heliosphere.
Measurements of the \fe60 pulse show that supernova radioisotopes rained on Earth for a long time, $\gtrsim 1 \ \rm Myr$.  The \fe60 flux is significantly longer than the gas profile of $\sim$~100 kyr \citep[][also seen in Fig. \ref{fig:Rstag}]{ertel_supernova_2022}: \fe60-bearing dust must decouple from the gas.
\citet{fry_magnetic_2020} built a model for dust propagation in an evolving supernova remnant that naturally reproduced the 1 Myr timescale.
Similarly, \citet{Slavin2020} and \citet{Sarangi2022} evolve models of supernova-formed dust with various compositions, though not focusing on \fe60.
Supernova dust models continue to be relevant for the present heliosphere because of the  ongoing infall of \fe60-bearing dust today \citep{koll_interstellar_2019}.

The continued infall of \fe60 onto Earth today points to the presence of \fe60-bearing dust in the very local ISM and its successful propagation to 1 au.  This raises questions that link heliophysics and astrophysics and requires further study.  
Initial models of supernova dust grain propagation into the magnetized solar wind were promising \citep{athanassiadou2011,fry2016}, but were simplistic in their treatment of the dust and of the solar wind.   A more complete picture of dust transport through the heliosphere is necessary for the last stage of the dust's journey \citep{linde2000, mann2010}.  Closely linked are issues of the astrophysics of supernova dust:
its production and survival
\citep{andersen2011,kirchschlager2020,kirchschlager2019,Slavin2020}, and the composition and sizes of \fe60-bearing grains \citep{Slavin2020_60Fe,fry_astrophysical_2015}.

In a different approach, \citet{opher_climate_2022} simulated the heliosphere in the presence of a cold cloud, showing that the heliosphere could shrink to a paltry 0.22 au. If supernovae had seeded the cloud with \fe60, the \fe60 would not have to be a component of dust in order to reach Earth.

If a supernova explodes too close to our solar system, it would be potentially harmful to life.
X-rays and cosmic rays (CRs) weaken the ozone layer, allowing solar UV radiation to cause significant biological damage \citep{ruderman_possible_1974, ellis_could_1995, gehrels_ozone_2003}.
The canonical ``kill distance'' is approximately 8 pc \citep{gehrels_ozone_2003}.
However, more distant supernovae may also have have severe consequences. 
\citet{fields_supernova_2020} postulated that a supernova about $\sim$ 20 pc away triggered the mass extinction in the end-Devonian 360 Myr ago.
Even at 100 pc, some biological damage may still occur \citep{thomas_terrestrial_2016, melott_hypothesis_2018}. The time period around 3 Myr ago was a critical period for human evolution, and astronomical factors may have played a role in this process \citep{melott_cosmic_2019, opher_climate_2022}.

Looking outwards, supernova remnants also provide an intriguing new interpretation for astrospheres.
Currently, surveys have found bow shocks of astrospheres around high-velocity stars \citep{peri_e-boss_2012} and in dusty regions like the Galactic plane \citep{kobulnicky_comprehensive_2016}.
The observation of astrospheres in SNRs would provide a novel form of stellar-interstellar interaction.

\section*{Opportunities for Heliophysics}

Motivated by abundant evidence for near-Earth supernova activity, we can re-contextualize many of the present-day heliosphere questions with the supernova-triggered Local Bubble replacing the a generic picture of the local ISM. 
Of the many questions this raises, 
some with direct impact on heliophysics are as follows.

{\bf How does supernova-forged dust propagate through the solar system to Earth, and how does this change as the SNR evolves and weakens?}
The detection of \fe60  in modern Antarctic snow \citep{koll_interstellar_2019} shows that supernova dust continues to enter the heliosphere and propagate to Earth.  The time is ripe for a careful theoretical study of the dust trajectory in the magnetized solar wind, to be tested against experimental searches for supernova-like interstellar dust in the solar system.   Such work would be guided by studies of interstellar dust trajectories in the present heliosphere, which reveal a rich dependence on grain size, charge-to-mass ratio, and solar wind polarity \citep{Slavin2012,sterken_flow_2012,Alexashov2016,Kruger2019}.  This work is directly relevant for the present-day \fe60 influx, which is arriving from the Local Interstellar Cloud.  But the material at early times is not from this cloud, and would arrive on supernova grains with speeds and properties that could be quite different from the present.

{\bf Given that SNRs produce Galactic cosmic rays, what are the consequences of the injection of freshly accelerated CRs with a different spectrum at the heliosphere boundary?}
It is an open question how cosmic rays propagate in the compressed heliosphere, and how this changes as the supernova blast fades. The interplay of the high cosmic ray flux on the solar wind is not known.  The particles will have higher energies than at present, and could lead to signatures in natural archives on the Earth.  Such explorations will also shed new light on models of CR acceleration in supernovae and their escape into the surrounding medium \citep{Blasi2013,Drury2011,Vieu2022,Ellison2000}.

{\bf  What signatures of recent near-Earth supernovae persist in the very local interstellar medium surrounding the heliosphere?}  What are predictions for {\em in situ} measurements of this environment, and how will such data shed light on the heliosphere's past?
Interstellar missions like Voyager and the proposed Interstellar Probe \citep{brandt_interstellar_2022, mcnutt_interstellar_2022} will not sample the average Galactic interstellar medium, but a part of the Local Bubble sculpted by the remains of ancient supernovae. Everything within the Local Bubble –- including the Local Interstellar Cloud –- has been processed by multiple supernova explosions millions of years ago. By examining the interstellar medium untainted by our heliosphere, Interstellar Probe may even contribute to our understanding of how old SNRs evolve and fade away into the Galaxy. 

{\bf Beyond \fe60-bearing dust, what other effects could supernovae have on Earth and its biosphere? Could evidence of these traces remain after 3 Myr?}  The expected high-energy and high-intensity cosmic-ray flux could lead to traces in the geological record, e.g., in the oldest ice cores and in deep-sea deposits. Such signatures would help to test ideas about biological consequences and biosphere damage.
As an extension, it is also interesting to probe the effects of supernovae on the Moon, the outer planets and other bodies that were exposed to the blast.
    
{\bf What can we learn from observing astrospheres of other stars and comparing with the heliosphere?} Extrasolar astropheres provide a testbed for stellar wind-ISM collisions under conditions different from our own, testing our understanding and possibly calibrating our models. These may also be probed by studies of the
effects of interstellar dust impacting interplanetary dust\footnote{C.~Lisse, {\it in preparation}.}.

\section*{Recommendations}

To realize fully the wide scientific opportunities we have outlined requires a wide range of experimental, observational, and theoretical work.  Our specific recommendations are as follows:

\begin{itemize}

\item {\bf Strengthen heliophysics links
to astrophysics and related fields
by encouraging and supporting interdisciplinary research.}  Cross-disciplinary research traditionally faces barriers to funding, with funding agents within each discipline often abdicating responsibility to others.  Encourage and support interdisciplinary collaboration, workshops, and public outreach.

\item {\bf Support present and future missions to the outer heliosphere and beyond, particularly the proposed Interstellar Probe.}  Such a mission would provide unique insight via direct detailed measurements of the local interstellar environment, including  gas and dust dynamical properties and composition, and magnetic fields.

\item  {\bf Support present and future missions that perform empirical dust measurements spanning 1 au to the outer solar system and that can identify and characterize interstellar grains.}  This would ground our understanding of propagation  of interstellar and supernova dust in the heliosphere.

\item 
{\bf Support theoretical study of the heliosphere over time, particularly in concert with the formation and evolution of the Local Bubble and the deposition of supernova-produced radioisotopes.}  Address questions we have outlined, including the propagation of supernova dust, supernova-accelerated cosmic rays,
the impact on solar system bodies, and signatures of these processes in the present geological and lunar records and across the solar system.
    
\item {\bf Support study of astrospheres of other stars as laboratories of heliophysics under a wide range of conditions.}  This can provide illuminating comparisons with the heliosphere today and in the presence of supernovae.
    
\end{itemize}

Fortunately, many of these efforts, including outer heliosphere missions such as Interstellar Probe, heliospheric dust measurements and cosmic-ray studies, are central to heliophysics writ large.  In addition, we believe that the interdisciplinary approach we advocate also strengthens heliophysics, raising its visibility to other sciences and to the public generally.  

The coming decade promises to offer exciting insights into our place in the evolving cosmos.  
This history is encoded in signatures from deep-sea deposits to the solar system to the stars.  We look forward to decoding these signals to unveil the dynamic nature of the heliosphere over time.

\pagebreak
\bibliography{references.bib}

\begin{thebibliography}{}
\expandafter\ifx\csname natexlab\endcsname\relax\def\natexlab#1{#1}\fi
\providecommand{\url}[1]{\href{#1}{#1}}
\providecommand{\dodoi}[1]{doi:~\href{http://doi.org/#1}{\nolinkurl{#1}}}
\providecommand{\doeprint}[1]{\href{http://ascl.net/#1}{\nolinkurl{http://ascl.net/#1}}}
\providecommand{\doarXiv}[1]{\href{https://arxiv.org/abs/#1}{\nolinkurl{https://arxiv.org/abs/#1}}}

\bibitem[{{Alexashov} {et~al.}(2016){Alexashov}, {Katushkina}, {Izmodenov}, \&
  {Akaev}}]{Alexashov2016}
{Alexashov}, D.~B., {Katushkina}, O.~A., {Izmodenov}, V.~V., \& {Akaev}, P.~S.
  2016, \mnras, 458, 2553, \dodoi{10.1093/mnras/stw514}

\bibitem[{Andersen {et~al.}(2011)Andersen, Rho, Reach, Hewitt, \&
  Bernard}]{andersen2011}
Andersen, M., Rho, J., Reach, W.~T., Hewitt, J.~W., \& Bernard, J.~P. 2011, The
  Astrophysical Journal, 742, 7, \dodoi{10.1088/0004-637X/742/1/7}

\bibitem[{Athanassiadou \& Fields(2011)}]{athanassiadou2011}
Athanassiadou, T., \& Fields, B.~D. 2011, New Astronomy, 16, 229,
  \dodoi{10.1016/j.newast.2010.09.007}

\bibitem[{Baranov \& Malama(1993)}]{baranov_model_1993}
Baranov, V.~B., \& Malama, Y.~G. 1993, J. Geophys. Res., 98, 15157,
  \dodoi{10.1029/93JA01171}

\bibitem[{{Ben{\'\i}tez} {et~al.}(2002){Ben{\'\i}tez},
  {Ma{\'\i}z-Apell{\'a}niz}, \& {Canelles}}]{Benitez2002}
{Ben{\'\i}tez}, N., {Ma{\'\i}z-Apell{\'a}niz}, J., \& {Canelles}, M. 2002,
  \prl, 88, 081101, \dodoi{10.1103/PhysRevLett.88.081101}

\bibitem[{Binns {et~al.}(2016)Binns, Israel, Christian, Cummings, de~Nolfo,
  Lave, Leske, Mewaldt, Stone, von Rosenvinge, \&
  Wiedenbeck}]{binns_observation_2016}
Binns, W.~R., Israel, M.~H., Christian, E.~R., {et~al.} 2016, Science, 352,
  677, \dodoi{10.1126/science.aad6004}

\bibitem[{{Blasi}(2013)}]{Blasi2013}
{Blasi}, P. 2013, \aapr, 21, 70, \dodoi{10.1007/s00159-013-0070-7}

\bibitem[{{Boschini} {et~al.}(2021){Boschini}, {Della Torre}, {Gervasi},
  {Grandi}, {J{\'o}hannesson}, {La Vacca}, {Masi}, {Moskalenko}, {Pensotti},
  {Porter}, {Quadrani}, {Rancoita}, {Rozza}, \& {Tacconi}}]{Boschini2021}
{Boschini}, M.~J., {Della Torre}, S., {Gervasi}, M., {et~al.} 2021, \apj, 913,
  5, \dodoi{10.3847/1538-4357/abf11c}

\bibitem[{{Branch} \& {Wheeler}(2017)}]{Branch2017}
{Branch}, D., \& {Wheeler}, J.~C. 2017, {Supernova Explosions} ({Springer}),
  \dodoi{10.1007/978-3-662-55054-0}

\bibitem[{Brandt {et~al.}(2022)Brandt, Provornikova, Cocoros, Turner,
  DeMajistre, Runyon, Lisse, Bale, Kurth, Galli, Wurz, McNutt,
  {Wimmer-Schweingruber}, Linsky, Redfield, Kollmann, Mandt, Rymer, Roelof,
  Kinnison, Opher, Hill, \& Paul}]{brandt_interstellar_2022}
Brandt, P.~C., Provornikova, E.~A., Cocoros, A., {et~al.} 2022, Acta
  Astronautica, 199, 364, \dodoi{10.1016/j.actaastro.2022.07.011}

\bibitem[{Breitschwerdt {et~al.}(2016)Breitschwerdt, Feige, Schulreich,
  de~Avillez, Dettbarn, \& Fuchs}]{breitschwerdt_locations_2016}
Breitschwerdt, D., Feige, J., Schulreich, M.~M., {et~al.} 2016, Nature, 532,
  73, \dodoi{10.1038/nature17424}

\bibitem[{Chaikin {et~al.}(2022)Chaikin, Kaurov, Fields, \&
  Correa}]{chaikin_simulations_2022}
Chaikin, E., Kaurov, A.~A., Fields, B.~D., \& Correa, C.~A. 2022, Monthly
  Notices of the Royal Astronomical Society, 512, 712,
  \dodoi{10.1093/mnras/stac327}

\bibitem[{{Drury}(2011)}]{Drury2011}
{Drury}, L.~O. 2011, \mnras, 415, 1807,
  \dodoi{10.1111/j.1365-2966.2011.18824.x}

\bibitem[{{Duch{\^e}ne} \& {Kraus}(2013)}]{Duchene2013}
{Duch{\^e}ne}, G., \& {Kraus}, A. 2013, \araa, 51, 269,
  \dodoi{10.1146/annurev-astro-081710-102602}

\bibitem[{Ellis \& Schramm(1995)}]{ellis_could_1995}
Ellis, J., \& Schramm, D.~N. 1995, PNAS, 92, 235, \dodoi{10.1073/pnas.92.1.235}

\bibitem[{Ellis {et~al.}(1996)Ellis, Fields, \& Schramm}]{Ellis:1995qb}
Ellis, J.~R., Fields, B.~D., \& Schramm, D.~N. 1996, Astrophys. J., 470, 1227,
  \dodoi{10.1086/177945}

\bibitem[{{Ellison} {et~al.}(2000){Ellison}, {Berezhko}, \&
  {Baring}}]{Ellison2000}
{Ellison}, D.~C., {Berezhko}, E.~G., \& {Baring}, M.~G. 2000, \apj, 540, 292,
  \dodoi{10.1086/309324}

\bibitem[{Ertel {et~al.}(2022)Ertel, Fry, Fields, \&
  Ellis}]{ertel_supernova_2022}
Ertel, A.~F., Fry, B.~J., Fields, B.~D., \& Ellis, J. 2022, arXiv:2206.06464
  [astro-ph], \dodoi{10.48550/arXiv.2206.06464}

\bibitem[{{Farhang} {et~al.}(2019){Farhang}, {van Loon}, {Khosroshahi},
  {Javadi}, \& {Bailey}}]{Farhang2019}
{Farhang}, A., {van Loon}, J.~T., {Khosroshahi}, H.~G., {Javadi}, A., \&
  {Bailey}, M. 2019, Nature Astronomy, 3, 922,
  \dodoi{10.1038/s41550-019-0814-z}

\bibitem[{Fields {et~al.}(2008)Fields, Athanassiadou, \&
  Johnson}]{fields_supernova_2008}
Fields, B.~D., Athanassiadou, T., \& Johnson, S.~R. 2008, ApJ, 678, 549,
  \dodoi{10.1086/523622}

\bibitem[{Fields {et~al.}(2020)Fields, Melott, Ellis, Ertel, Fry, Lieberman,
  Liu, Miller, \& Thomas}]{fields_supernova_2020}
Fields, B.~D., Melott, A.~L., Ellis, J., {et~al.} 2020, PNAS, 117, 21008,
  \dodoi{10.1073/pnas.2013774117}

\bibitem[{Fimiani {et~al.}(2016)Fimiani, Cook, Faestermann,
  {G{\'o}mez-Guzm{\'a}n}, Hain, Herzog, Knie, Korschinek, Ludwig, Park, Reedy,
  \& Rugel}]{fimiani_interstellar_2016}
Fimiani, L., Cook, D.~L., Faestermann, T., {et~al.} 2016, Physical Review
  Letters, 116, \dodoi{10.1103/PhysRevLett.116.151104}

\bibitem[{Fitoussi {et~al.}(2008)Fitoussi, Raisbeck, Knie, Korschinek,
  Faestermann, Goriely, Lunney, Poutivtsev, Rugel, Waelbroeck, \&
  Wallner}]{fitoussi_search_2008}
Fitoussi, C., Raisbeck, G.~M., Knie, K., {et~al.} 2008, Phys. Rev. Lett., 101,
  121101, \dodoi{10.1103/PhysRevLett.101.121101}

\bibitem[{Frisch \& Dwarkadas(2017)}]{frisch_effect_2017}
Frisch, P., \& Dwarkadas, V.~V. 2017, in Handbook of {{Supernovae}}, ed. A.~W.
  Alsabti \& P.~Murdin ({Cham}: {Springer International Publishing}),
  2253--2285, \dodoi{10.1007/978-3-319-21846-5-13}

\bibitem[{Frisch {et~al.}(2011)Frisch, Redfield, \&
  Slavin}]{frisch_interstellar_2011}
Frisch, P.~C., Redfield, S., \& Slavin, J.~D. 2011, Annual Review of Astronomy
  and Astrophysics, 49, 237, \dodoi{10.1146/annurev-astro-081710-102613}

\bibitem[{Fry {et~al.}(2015)Fry, Fields, \& Ellis}]{fry_astrophysical_2015}
Fry, B.~J., Fields, B.~D., \& Ellis, J.~R. 2015, The Astrophysical Journal,
  800, 71, \dodoi{10.1088/0004-637X/800/1/71}

\bibitem[{Fry {et~al.}(2016)Fry, Fields, \& Ellis}]{fry2016}
---. 2016, The Astrophysical Journal, 827, 48,
  \dodoi{10.3847/0004-637X/827/1/48}

\bibitem[{Fry {et~al.}(2020)Fry, Fields, \& Ellis}]{fry_magnetic_2020}
---. 2020, ApJ, 894, 109, \dodoi{10.3847/1538-4357/ab86bf}

\bibitem[{{Galeazzi} {et~al.}(2014){Galeazzi}, {Chiao}, {Collier}, {Cravens},
  {Koutroumpa}, {Kuntz}, {Lallement}, {Lepri}, {McCammon}, {Morgan}, {Porter},
  {Robertson}, {Snowden}, {Thomas}, {Uprety}, {Ursino}, \&
  {Walsh}}]{Galeazzi2014}
{Galeazzi}, M., {Chiao}, M., {Collier}, M.~R., {et~al.} 2014, \nat, 512, 171,
  \dodoi{10.1038/nature13525}

\bibitem[{Gehrels {et~al.}(2003)Gehrels, Laird, Jackman, Cannizzo, Mattson, \&
  Chen}]{gehrels_ozone_2003}
Gehrels, N., Laird, C.~M., Jackman, C.~H., {et~al.} 2003, ApJ, 585, 1169,
  \dodoi{10.1086/346127}

\bibitem[{{Gry} \& {Jenkins}(2014)}]{Gry2014}
{Gry}, C., \& {Jenkins}, E.~B. 2014, \aap, 567, A58,
  \dodoi{10.1051/0004-6361/201323342}

\bibitem[{Kachelrie{\ss} {et~al.}(2015)Kachelrie{\ss}, Neronov, \&
  Semikoz}]{kachelries_signatures_2015}
Kachelrie{\ss}, M., Neronov, A., \& Semikoz, D.~V. 2015, Physical Review
  Letters, 115, 181103, \dodoi{10.1103/PhysRevLett.115.181103}

\bibitem[{Kachelrie{\ss} {et~al.}(2018)Kachelrie{\ss}, Neronov, \&
  Semikoz}]{kachelries_cosmic_2018}
---. 2018, Phys. Rev. D, 97, 063011, \dodoi{10.1103/PhysRevD.97.063011}

\bibitem[{Kirchschlager {et~al.}(2020)Kirchschlager, Barlow, \&
  Schmidt}]{kirchschlager2020}
Kirchschlager, F., Barlow, M.~J., \& Schmidt, F.~D. 2020, The Astrophysical
  Journal, 893, 70, \dodoi{10.3847/1538-4357/ab7db8}

\bibitem[{Kirchschlager {et~al.}(2019)Kirchschlager, Schmidt, Barlow, Fogerty,
  Bevan, \& Priestley}]{kirchschlager2019}
Kirchschlager, F., Schmidt, F.~D., Barlow, M.~J., {et~al.} 2019, Monthly
  Notices of the Royal Astronomical Society, 489, 4465,
  \dodoi{10.1093/mnras/stz2399}

\bibitem[{Knie {et~al.}(2004)Knie, Korschinek, Faestermann, Dorfi, Rugel, \&
  Wallner}]{knie_60fe_2004}
Knie, K., Korschinek, G., Faestermann, T., {et~al.} 2004, Physical Review
  Letters, 93, \dodoi{10.1103/PhysRevLett.93.171103}

\bibitem[{Knie {et~al.}(1999)Knie, Merchel, Korschinek, Faestermann, Herpers,
  Gloris, \& Michel}]{knie_ams_1999}
Knie, K., Merchel, S., Korschinek, G., {et~al.} 1999, Meteoritics and Planetary
  Science, 34, 729, \dodoi{10.1111/j.1945-5100.1999.tb01385.x}

\bibitem[{Kobulnicky {et~al.}(2016)Kobulnicky, Chick, Schurhammer, Andrews,
  Povich, Munari, Olivier, Sorber, Wernke, Dale, \&
  Dixon}]{kobulnicky_comprehensive_2016}
Kobulnicky, H.~A., Chick, W.~T., Schurhammer, D.~P., {et~al.} 2016, ApJS, 227,
  18, \dodoi{10.3847/0067-0049/227/2/18}

\bibitem[{Koll {et~al.}(2019)Koll, Korschinek, Faestermann,
  {G{\'o}mez-Guzm{\'a}n}, Kipfstuhl, Merchel, \&
  Welch}]{koll_interstellar_2019}
Koll, D., Korschinek, G., Faestermann, T., {et~al.} 2019, Phys. Rev. Lett.,
  123, 072701, \dodoi{10.1103/PhysRevLett.123.072701}

\bibitem[{Korschinek {et~al.}(2020)Korschinek, Faestermann, Poutivtsev, Arazi,
  Knie, Rugel, \& Wallner}]{korschinek_supernova-produced_2020}
Korschinek, G., Faestermann, T., Poutivtsev, M., {et~al.} 2020, Phys. Rev.
  Lett., 125, 031101, \dodoi{10.1103/PhysRevLett.125.031101}

\bibitem[{{Kr{\"u}ger} {et~al.}(2019){Kr{\"u}ger}, {Strub}, {Altobelli},
  {Sterken}, {Srama}, \& {Gr{\"u}n}}]{Kruger2019}
{Kr{\"u}ger}, H., {Strub}, P., {Altobelli}, N., {et~al.} 2019, \aap, 626, A37,
  \dodoi{10.1051/0004-6361/201834316}

\bibitem[{{Lallement} {et~al.}(2019){Lallement}, {Babusiaux}, {Vergely},
  {Katz}, {Arenou}, {Valette}, {Hottier}, \& {Capitanio}}]{Lallement2019}
{Lallement}, R., {Babusiaux}, C., {Vergely}, J.~L., {et~al.} 2019, \aap, 625,
  A135, \dodoi{10.1051/0004-6361/201834695}

\bibitem[{Linde \& Gombosi(2000)}]{linde2000}
Linde, T.~J., \& Gombosi, T.~I. 2000, Journal of Geophysical Research: Space
  Physics, 105, 10411, \dodoi{10.1029/1999JA900149}

\bibitem[{Ludwig {et~al.}(2016)Ludwig, Bishop, Egli, Chernenko, Deneva,
  Faestermann, Famulok, Fimiani, {G{\'o}mez-Guzm{\'a}n}, Hain, Korschinek,
  Hanzlik, Merchel, \& Rugel}]{ludwig_time-resolved_2016}
Ludwig, P., Bishop, S., Egli, R., {et~al.} 2016, Proceedings of the National
  Academy of Sciences, 113, 9232, \dodoi{10.1073/pnas.1601040113}

\bibitem[{{Ma{\'\i}z-Apell{\'a}niz}(2001)}]{Maiz-Apellaniz_2001}
{Ma{\'\i}z-Apell{\'a}niz}, J. 2001, ApJL, 560, L83, \dodoi{10.1086/324016}

\bibitem[{{Mamajek}(2016)}]{Mamajek2016}
{Mamajek}, E.~E. 2016, in Young Stars \& Planets Near the Sun, ed. J.~H.
  {Kastner}, B.~{Stelzer}, \& S.~A. {Metchev}, Vol. 314, 21--26,
  \dodoi{10.1017/S1743921315006250}

\bibitem[{Mann(2010)}]{mann2010}
Mann, I. 2010, Annual Review of Astronomy and Astrophysics, 48, 173,
  \dodoi{10.1146/annurev-astro-081309-130846}

\bibitem[{McNutt {et~al.}(2022)McNutt, {Wimmer-Schweingruber}, Gruntman,
  Krimigis, Roelof, Brandt, Vernon, Paul, Stough, \&
  Kinnison}]{mcnutt_interstellar_2022}
McNutt, R.~L., {Wimmer-Schweingruber}, R.~F., Gruntman, M., {et~al.} 2022, Acta
  Astronautica, 196, 13, \dodoi{10.1016/j.actaastro.2022.04.001}

\bibitem[{Melott {et~al.}(2018)Melott, Marinho, \&
  Paulucci}]{melott_hypothesis_2018}
Melott, A.~L., Marinho, F., \& Paulucci, L. 2018, Astrobiology, 19, 825,
  \dodoi{10.1089/ast.2018.1902}

\bibitem[{Melott \& Thomas(2019)}]{melott_cosmic_2019}
Melott, A.~L., \& Thomas, B.~C. 2019, The Journal of Geology, 127, 475,
  \dodoi{10.1086/703418}

\bibitem[{Miller \& Fields(2022)}]{miller_heliospheric_2022}
Miller, J.~A., \& Fields, B.~D. 2022, The Astrophysical Journal, 934, 32,
  \dodoi{10.3847/1538-4357/ac77f1}

\bibitem[{{Motte} {et~al.}(2018){Motte}, {Bontemps}, \& {Louvet}}]{Motte2018}
{Motte}, F., {Bontemps}, S., \& {Louvet}, F. 2018, \araa, 56, 41,
  \dodoi{10.1146/annurev-astro-091916-055235}

\bibitem[{M{\"u}ller {et~al.}(2009)M{\"u}ller, Frisch, Fields, \&
  Zank}]{muller_heliosphere_2009}
M{\"u}ller, H.-R., Frisch, P.~C., Fields, B.~D., \& Zank, G.~P. 2009, Space Sci
  Rev, 143, 415, \dodoi{10.1007/s11214-008-9448-7}

\bibitem[{M{\"u}ller {et~al.}(2006)M{\"u}ller, Frisch, Florinski, \&
  Zank}]{muller_heliospheric_2006}
M{\"u}ller, H.-R., Frisch, P.~C., Florinski, V., \& Zank, G.~P. 2006, ApJ, 647,
  1491, \dodoi{10.1086/505588}

\bibitem[{Neuh{\"a}user {et~al.}(2020)Neuh{\"a}user, Gie{\ss}ler, \&
  Hambaryan}]{neuhauser_nearby_2020}
Neuh{\"a}user, R., Gie{\ss}ler, F., \& Hambaryan, V.~V. 2020, Monthly Notices
  of the Royal Astronomical Society, 498, 899, \dodoi{10.1093/mnras/stz2629}

\bibitem[{Opher \& Loeb(2022)}]{opher_climate_2022}
Opher, M., \& Loeb, A. 2022, arXiv:2202.01813 [astro-ph].
\newblock \doarXiv{2202.01813}

\bibitem[{Peri {et~al.}(2012)Peri, Benaglia, Brookes, Stevens, \&
  Isequilla}]{peri_e-boss_2012}
Peri, C.~S., Benaglia, P., Brookes, D.~P., Stevens, I.~R., \& Isequilla, N.~L.
  2012, A\&A, 538, A108, \dodoi{10.1051/0004-6361/201118116}

\bibitem[{Ruderman(1974)}]{ruderman_possible_1974}
Ruderman, M.~A. 1974, Science, 184, 1079, \dodoi{10.1126/science.184.4141.1079}

\bibitem[{{Sarangi} \& {Slavin}(2022)}]{Sarangi2022}
{Sarangi}, A., \& {Slavin}, J.~D. 2022, \apj, 933, 89,
  \dodoi{10.3847/1538-4357/ac713d}

\bibitem[{Savchenko {et~al.}(2015)Savchenko, Kachelrie{\ss}, \&
  Semikoz}]{savchenko_imprint_2015}
Savchenko, V., Kachelrie{\ss}, M., \& Semikoz, D.~V. 2015, The Astrophysical
  Journal, 809, L23, \dodoi{10.1088/2041-8205/809/2/L23}

\bibitem[{Schulreich {et~al.}(2017)Schulreich, Breitschwerdt, Feige, \&
  Dettbarn}]{schulreich_numerical_2017}
Schulreich, M.~M., Breitschwerdt, D., Feige, J., \& Dettbarn, C. 2017, A\&A,
  604, A81, \dodoi{10.1051/0004-6361/201629837}

\bibitem[{{Slavin}(2020)}]{Slavin2020_60Fe}
{Slavin}, J. 2020, in Journal of Physics Conference Series, Vol. 1620, Journal
  of Physics Conference Series, 012019, \dodoi{10.1088/1742-6596/1620/1/012019}

\bibitem[{{Slavin} {et~al.}(2020){Slavin}, {Dwek}, {Mac Low}, \&
  {Hill}}]{Slavin2020}
{Slavin}, J.~D., {Dwek}, E., {Mac Low}, M.-M., \& {Hill}, A.~S. 2020, \apj,
  902, 135, \dodoi{10.3847/1538-4357/abb5a4}

\bibitem[{{Slavin} {et~al.}(2012){Slavin}, {Frisch}, {M{\"u}ller},
  {Heerikhuisen}, {Pogorelov}, {Reach}, \& {Zank}}]{Slavin2012}
{Slavin}, J.~D., {Frisch}, P.~C., {M{\"u}ller}, H.-R., {et~al.} 2012, \apj,
  760, 46, \dodoi{10.1088/0004-637X/760/1/46}

\bibitem[{Smith \& Cox(2001)}]{smith_multiple_2001}
Smith, R.~K., \& Cox, D.~P. 2001, ApJS, 134, 283, \dodoi{10.1086/320850}

\bibitem[{{Snowden} {et~al.}(2015){Snowden}, {Heiles}, {Koutroumpa}, {Kuntz},
  {Lallement}, {McCammon}, \& {Peek}}]{Snowden2015}
{Snowden}, S.~L., {Heiles}, C., {Koutroumpa}, D., {et~al.} 2015, \apj, 806,
  119, \dodoi{10.1088/0004-637X/806/1/119}

\bibitem[{Sok{\'o}{\l} {et~al.}(2019)Sok{\'o}{\l}, Kubiak, \&
  Bzowski}]{sokol_interstellar_2019}
Sok{\'o}{\l}, J.~M., Kubiak, M.~A., \& Bzowski, M. 2019, The Astrophysical
  Journal, 879, 24, \dodoi{10.3847/1538-4357/ab21c4}

\bibitem[{{Sok{\'o}{\l}} {et~al.}(2022){Sok{\'o}{\l}}, {Kucharek}, {Baliukin},
  {Fahr}, {Izmodenov}, {Kornbleuth}, {Mostafavi}, {Opher}, {Park}, {Pogorelov},
  {Quinn}, {Smith}, {Zank}, \& {Zhang}}]{Sokol2022}
{Sok{\'o}{\l}}, J.~M., {Kucharek}, H., {Baliukin}, I.~I., {et~al.} 2022, \ssr,
  218, 18, \dodoi{10.1007/s11214-022-00883-6}

\bibitem[{Sterken {et~al.}(2012)Sterken, Altobelli, Kempf, Schwehm, Srama, \&
  Gr{\"u}n}]{sterken_flow_2012}
Sterken, V.~J., Altobelli, N., Kempf, S., {et~al.} 2012, Astronomy \&
  Astrophysics, 538, A102, \dodoi{10.1051/0004-6361/201117119}

\bibitem[{Thomas {et~al.}(2016)Thomas, Engler, Kachelrie{\ss}, Melott,
  Overholt, \& Semikoz}]{thomas_terrestrial_2016}
Thomas, B.~C., Engler, E.~E., Kachelrie{\ss}, M., {et~al.} 2016, The
  Astrophysical Journal, 826, L3, \dodoi{10.3847/2041-8205/826/1/L3}

\bibitem[{{Vieu} {et~al.}(2022){Vieu}, {Gabici}, {Tatischeff}, \&
  {Ravikularaman}}]{Vieu2022}
{Vieu}, T., {Gabici}, S., {Tatischeff}, V., \& {Ravikularaman}, S. 2022,
  \mnras, 512, 1275, \dodoi{10.1093/mnras/stac543}

\bibitem[{Wallner {et~al.}(2016)Wallner, Feige, Kinoshita, Paul, Fifield,
  Golser, Honda, Linnemann, Matsuzaki, Merchel, Rugel, Tims, Steier, Yamagata,
  \& Winkler}]{wallner_recent_2016}
Wallner, A., Feige, J., Kinoshita, N., {et~al.} 2016, Nature, 532, 69,
  \dodoi{10.1038/nature17196}

\bibitem[{Wallner {et~al.}(2021)Wallner, Froehlich, Hotchkis, Kinoshita, Paul,
  Martschini, Pavetich, Tims, Kivel, Schumann, Honda, Matsuzaki, \&
  Yamagata}]{wallner_60fe_2021}
Wallner, A., Froehlich, M.~B., Hotchkis, M. a.~C., {et~al.} 2021, Science, 372,
  742, \dodoi{10.1126/science.aax3972}

\bibitem[{{Welsh} {et~al.}(2010){Welsh}, {Lallement}, {Vergely}, \&
  {Raimond}}]{Welsh_etal_2010}
{Welsh}, B.~Y., {Lallement}, R., {Vergely}, J.~L., \& {Raimond}, S. 2010, \aap,
  510, A54, \dodoi{10.1051/0004-6361/200913202}

\bibitem[{Zank(1999)}]{zank_interaction_1999}
Zank, G.~P. 1999, Space Sci Rev, 89, 413, \dodoi{10.1023/A:1005155601277}

\bibitem[{{Zucker} {et~al.}(2022){Zucker}, {Goodman}, {Alves}, {Bialy},
  {Foley}, {Speagle}, {Gro{\^I}{\texttwosuperior}schedl}, {Finkbeiner},
  {Burkert}, {Khimey}, \& {Swiggum}}]{Zucker2022}
{Zucker}, C., {Goodman}, A.~A., {Alves}, J., {et~al.} 2022, \nat, 601, 334,
  \dodoi{10.1038/s41586-021-04286-5}

\end{thebibliography}
\bibliographystyle{aasjournal}

\end{document}